\documentclass[lettersize,journal]{IEEEtran}
\usepackage{amsmath,amsfonts}
\usepackage{physics,amsmath}
\usepackage{algorithm}
\usepackage{algpseudocode}
\usepackage{multirow}
\usepackage{array}
\usepackage{bm}
\usepackage{placeins}
\usepackage {wrapfig}
\usepackage{xcolor}
\usepackage{tcolorbox}
\usepackage{amsfonts}

\usepackage{amsmath,amssymb}

\usepackage{subcaption}
\usepackage{mathtools}
\usepackage{xcolor}
\usepackage{float} 
\usepackage{textcomp}
\usepackage{stfloats}
\usepackage{amsmath}   

\usepackage{lipsum}  
\usepackage{verbatim}
\usepackage{graphicx}
\usepackage{pdfpages}
\usepackage{tikz}
\usepackage{lettrine}
\usetikzlibrary{3d}
\usetikzlibrary{arrows.meta} 

\usepackage[colorlinks=true, linkcolor=blue, urlcolor=blue, citecolor=blue, anchorcolor=blue]{hyperref}

\begin{document}

\title{Fundamental Limits of Cooperative Integrated Sensing and Communications over Low-Earth Orbit THz Satellite Channels}
\author{Haofan Dong,~\IEEEmembership{Member,~IEEE}, Houtianfu Wang,~\IEEEmembership{Student Member,~IEEE,} Hanlin Cai,~\IEEEmembership{Student Member,~IEEE,}
        and Ozgur B. Akan,~\IEEEmembership{Fellow,~IEEE}
\thanks{The authors are with Internet of Everything Group, Department of Engineering, University of Cambridge, CB3 0FA Cambridge, UK.}
\thanks{Ozgur B. Akan is also with the Center for neXt-generation Communications
(CXC), Department of Electrical and Electronics Engineering, Koç University, 34450 Istanbul, Turkey (email:oba21@cam.ac.uk)}}
	   



\maketitle

\begin{abstract}

Terahertz inter-satellite links enable  high-precision sensing for Low Earth Orbit (LEO) constellations, yet face fundamental bounds from hardware impairments, pointing errors, and network interference. We develop a   Network Cramér-Rao Lower Bound (N-CRLB) framework incorporating dynamic topology, hardware quality factor $\Gamma_{\text{eff}}$, phase noise $\sigma^2_\phi$, and cooperative effects through recursive Fisher Information analysis. Our analysis reveals three key insights: (i) hardware and phase noise create power-independent performance ceilings ($\sigma_{\text{ceiling}} \propto \sqrt{\Gamma_{\text{eff}}}$) and floors ($\sigma_{\text{floor}} \propto \sqrt{\sigma^2_\phi}/f_c$), with power-only scaling saturating above $\text{SNR}_{\text{crit}}=1/\Gamma_{\text{eff}}$; (ii) interference coefficients $\alpha_{\ell m}$ enable opportunistic sensing with demonstrated gains of 5.67~dB under specific conditions (65~dB processing gain, 50~dBi antennas); (iii) measurement correlations from shared timing references, when properly modeled, do not degrade performance and can provide common-mode rejection benefits compared to mismodeled independent-noise baselines. Sub-millimeter ranging requires co-optimized hardware ($\Gamma_{\text{eff}}<0.01$), oscillators ($\sigma^2_\phi<10^{-2}$), and appropriate 3D geometry configurations.
\end{abstract}


\section{Introduction}

Terahertz inter-satellite links promise to significantly advance Low Earth Orbit (LEO) mega-constellations, offering multi-terabit-per-second data rates through bandwidths exceeding 100 GHz \cite{chen2022tutorial}. The integration of sensing with these communication links—enabled by the sub-millimeter wavelengths at THz frequencies—could transform thousands of satellites into a distributed sensing network \cite{wu2025enhancing}. Yet scaling from single-link demonstrations to networked constellations reveals a fundamental paradox: the same physical characteristics enabling ISAC performance \cite{gu2024isac} become critical vulnerability points when hundreds of links operate concurrently \cite{niu2024interference}. This transition demands new theoretical foundations that capture emergent network phenomena absent in isolated links \cite{you2024ubiquitous}.

The transition from single-link demonstrations to networked THz-ISAC systems presents fundamental challenges requiring comprehensive theoretical treatment. When scaling to constellations with hundreds or thousands of nodes, three interrelated phenomena emerge as critical: THz signals exhibit high sensitivity to hardware impairments—phase noise scaling quadratically with carrier frequency and power amplifier nonlinearities creating signal-dependent distortions \cite{chen2022mcrb}. These effects compound with platform-induced pointing errors, where microradian-level deviations cause significant gain fluctuations \cite{dabiri2022pointing,zaman2020impact}. Additionally, the dynamic topology of LEO constellations, with relative velocities reaching 15 km/s, creates continuously evolving network structures that challenge traditional resource allocation \cite{gao2022research}. Dense frequency reuse generates complex interference patterns exhibiting intermittent coupling through pointing jitter rather than simple additive noise \cite{caceres2024interference,ruan2015generalized}.

While recent THz-ISAC frameworks have demonstrated debris detection \cite{dong2025debrisense} and planetary sensing capabilities \cite{dong2024martian}, existing work remains confined to single-link analysis that does not adequately capture networked phenomena. Our previous framework \cite{dong2025fundamental} revealed hardware-imposed capacity ceilings but overlooked critical network effects: how structured interference from concurrent transmissions creates information rather than noise \cite{vaghefi2015cooperative}, how measurement correlations from shared resources alter fundamental scaling laws \cite{ghasemi2010interference}, and how distributed sensing transforms unobservable states into jointly estimable quantities \cite{olfati2007consensus}. To the best of our knowledge, no existing framework comprehensively captures hardware physics, network topology, and information-theoretic limits for THz LEO-ISAC systems.

This paper establishes a comprehensive framework for characterizing performance limits of networked THz LEO-ISAC systems through three key contributions:

\begin{enumerate}
\item A recursive Network Cramér-Rao Lower Bound (N-CRLB) framework is developed that jointly models dynamic topology, hardware impairments (quality factor $\Gamma_{\text{eff}}$, phase noise $\sigma^2_\phi$), and structured interference through information filtering. Analysis reveals that sub-millimeter ranging precision requires hardware quality $\Gamma_{\text{eff}}<0.01$ and phase stability $\sigma^2_\phi<10^{-2}$, with performance saturating at power levels above $\text{SNR}_{\text{crit}}=1/\Gamma_{\text{eff}}$.

\item Closed-form expressions for interference coefficients $\alpha_{\ell m}$ are derived, quantifying conditions under which interference enables opportunistic sensing. Numerical validation demonstrates 5.67 dB information gain when processing gain exceeds 65 dB, establishing the crossover point where interference transitions from detrimental to beneficial.

\item The analysis addresses measurement correlations induced by shared timing references, demonstrating that proper modeling enables common-mode rejection while mischaracterization as independent noise degrades performance by up to 53\% at correlation coefficient $\rho>0.75$. This reveals fundamental scaling laws for cooperative sensing in correlated noise environments.
\end{enumerate}

The remainder of this paper is organized as follows: Section II establishes the networked THz ISL-ISAC system model; Section III develops the N-CRLB framework; Section IV analyzes interference and opportunistic sensing; Section V validates performance through numerical analysis; Section VI concludes.

\section{System Model}

\begin{figure}[!t]
    \centering
    \includegraphics[width=\columnwidth]{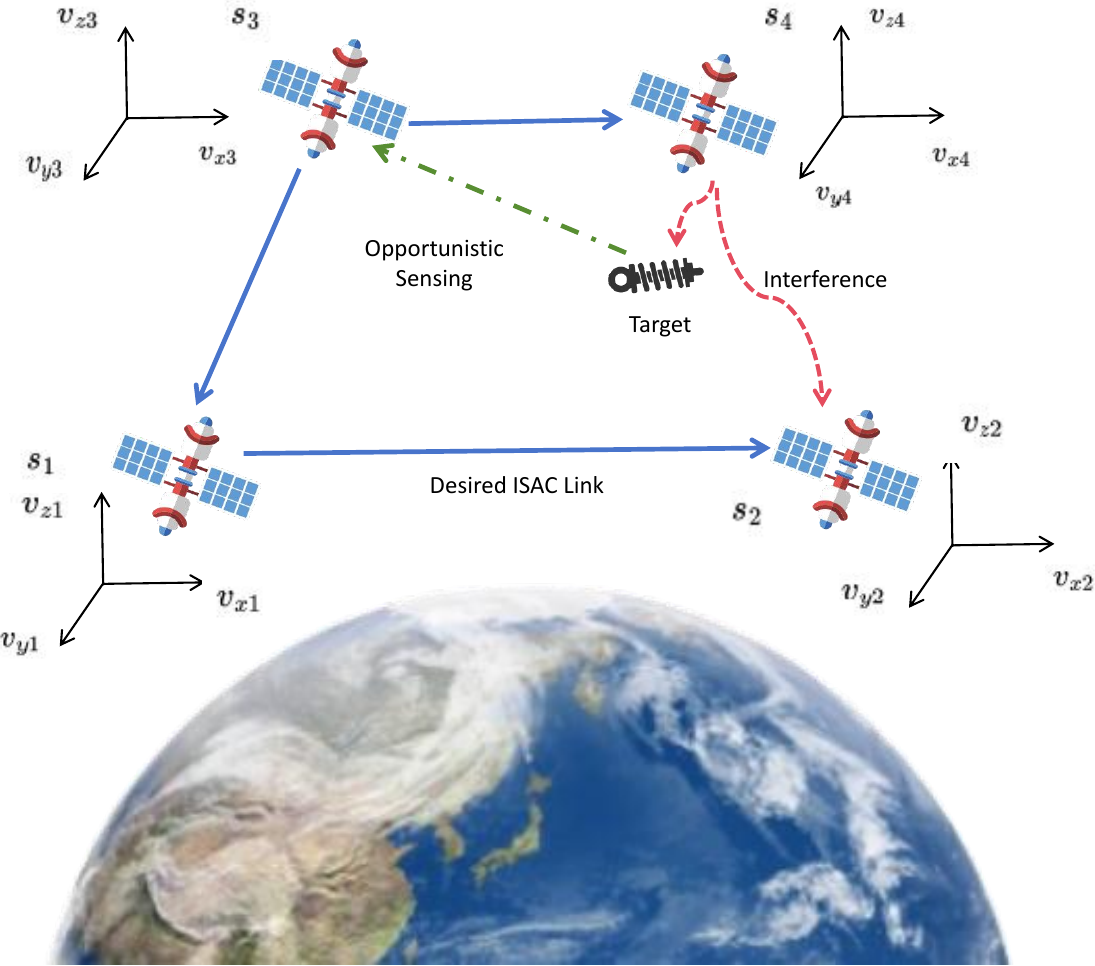}
    \caption{System model for a networked THz LEO-ISAC constellation. The framework captures desired ISAC links used for communication and cooperative self-positioning, inter-link interference, and the transformation of interference signals into an asset for opportunistic sensing of a non-cooperative target. The high-velocity dynamics of each satellite are explicitly considered.}
    \label{fig:system_model}
\end{figure}

The networked LEO-ISAC scenario under consideration is depicted in Fig.~\ref{fig:system_model}. The system comprises multiple high-velocity LEO satellites that form a dynamic communication and sensing network. Desired ISAC links are established for cooperative self-positioning and data exchange. However, in a spectrally crowded environment, these links are subject to inter-link interference. Our framework uniquely models this interference not merely as a detrimental factor, but also as a valuable resource. As shown, signals from an interfering satellite can illuminate a non-cooperative target, with the scattered echo received by another satellite, thus creating an opportunistic sensing pathway. This dual treatment of interference as both a challenge to communication and an asset for sensing is a central theme of our analysis.

\subsection{Network Topology and State Vector}

We model the LEO satellite constellation as a time-varying undirected graph $G(t)=(\mathcal{V}, \mathcal{E}(t))$, where $\mathcal{V} = \{v_1, v_2, \ldots, v_{N_v}\}$ denotes the set of $N_v$ satellite nodes and $\mathcal{E}(t) \subseteq \mathcal{V} \times \mathcal{V}$ represents active inter-satellite links (ISLs) evolving with orbital dynamics \cite{hong2024cooperative, shen2010fundamental}.

The network state $\mathbf{x}_k = [\mathbf{x}_{1,k}^T, \ldots, \mathbf{x}_{N_v,k}^T]^T \in \mathbb{R}^{8N_v}$ evolves according to $\mathbf{x}_k = \mathbf{F}_k \mathbf{x}_{k-1} + \mathbf{w}_{k-1}$, where each satellite state $\mathbf{x}_{v,k} = [\mathbf{p}_{v,k}^T, \dot{\mathbf{p}}_{v,k}^T, b_{v,k}, \dot{b}_{v,k}]^T$ includes position, velocity, and clock parameters\cite{krawinkel2016benefits}.

To achieve sub-millimeter precision required by THz ISAC, the state transition matrix incorporates Earth's J2 perturbation. The continuous dynamics in the J2000 Earth-Centered Inertial (ECI) frame are \cite{vess2003study}:
\begin{equation}
\ddot{\mathbf{p}} = -\frac{\mu}{p^3}\mathbf{p} + \mathbf{a}_{J_2}(\mathbf{p}), \quad \mathbf{a}_{J_2} = -\frac{3\mu J_2 R_e^2}{2p^5}\begin{bmatrix} x(1-5z^2/p^2) \\ y(1-5z^2/p^2) \\ z(3-5z^2/p^2) \end{bmatrix},
\end{equation}
where $J_2 = 1.08263 \times 10^{-3}$, $R_e = 6371$ km, and $\mu = 3.986 \times 10^{14}$ m$^3$/s$^2$.

The linearized system $\delta\dot{\mathbf{x}}(t) = \mathbf{A}(t)\delta\mathbf{x}(t)$ yields the discrete-time transition matrix through second-order approximation:
\begin{equation}
\mathbf{F}_k \approx \mathbf{I}_8 + \bar{\mathbf{A}}_k\Delta t + \frac{1}{2}\bar{\mathbf{A}}_k^2\Delta t^2, \quad \bar{\mathbf{A}}_k = \mathbf{A}(t_{k-1} + \Delta t/2),
\end{equation}
ensuring $O(\Delta t^3)$ local accuracy with sub-millimeter prediction error for $\Delta t \leq 1$ s.

To provide physical grounding, we adopt a velocity random walk (VRW) model for the process noise covariance:
\begin{align}
[\mathbf{Q}_k]_{\mathbf{p}\mathbf{p}} &= \sigma^2_a \frac{\Delta t^3}{3} \mathbf{I}_{3\times3}, \\
[\mathbf{Q}_k]_{\mathbf{p}\dot{\mathbf{p}}} &= \sigma^2_a \frac{\Delta t^2}{2} \mathbf{I}_{3\times3}, \\
[\mathbf{Q}_k]_{\dot{\mathbf{p}}\dot{\mathbf{p}}} &= \sigma^2_a \Delta t \mathbf{I}_{3\times3},
\end{align}
where $\sigma_a^2$ represents the power spectral density of acceleration noise. 
The explicit $J_2$ modeling in $\mathbf{F}_k$ captures the dominant systematic perturbation 
($2.5 \,\mathrm{mm}$ error at $10 \,\mathrm{s}$), while atmospheric drag
($< 0.05 \,\mathrm{mm}$), solar radiation pressure ($< 1 \,\mu\mathrm{m}$), 
and third-body effects ($< 1 \,\mu\mathrm{m}$) remain in stochastic process
noise $\mathbf{Q}_k$ due to their smaller magnitudes or variable nature. 
Clock noise follows Allan variance scaling with 
$[\mathbf{Q}_k]_{bb} = \sigma_y^2 c^2 \Delta t$.

\subsection{Multi-Link Measurement Model}

For an active ISL $\ell \in \mathcal{E}(t_k)$ connecting transmitting satellite $v_i$ and receiving satellite $v_j$, the Time of Arrival (TOA) measurement is:
\begin{equation}
z_{\ell,k} = h_\ell(\mathbf{x}_k) + v_{\ell,k},
\end{equation}
where the nonlinear measurement function captures geometric propagation delay and clock offset:
\begin{equation}
h_\ell(\mathbf{x}_k) = \frac{1}{c} \|\mathbf{p}_{j,k} - \mathbf{p}_{i,k}\| + (b_{j,k} - b_{i,k}),
\end{equation}

For Fisher Information Matrix construction, we require the measurement Jacobian. Given that $h_\ell(\mathbf{x}_k)$ depends only on satellites $v_i$ and $v_j$, the Jacobian row vector $\mathbf{H}_\ell = \nabla_{\mathbf{x}_k} h_\ell(\mathbf{x}_k) \in \mathbb{R}^{1 \times 8N_v}$ exhibits sparse structure with non-zero elements:
\begin{align}
\frac{\partial h_\ell}{\partial \mathbf{p}_{i,k}} &= -\frac{\mathbf{u}_{ij}^T}{c}, \quad \frac{\partial h_\ell}{\partial \mathbf{p}_{j,k}} = \frac{\mathbf{u}_{ij}^T}{c}, \\
\frac{\partial h_\ell}{\partial b_{i,k}} &= -1, \quad \frac{\partial h_\ell}{\partial b_{j,k}} = 1,
\end{align}
where $\mathbf{u}_{ij} = (\mathbf{p}_{j,k} - \mathbf{p}_{i,k})/\|\mathbf{p}_{j,k} - \mathbf{p}_{i,k}\|$ is the unit line-of-sight vector. All partial derivatives with respect to velocities and clock drifts vanish.

Each scalar TOA measurement constrains only one degree of freedom in the 16-dimensional joint state space of two satellites. Full network observability requires multiple ISLs with diverse geometric configurations, analogous to the Geometric Dilution of Precision (GDOP) concept in GNSS \cite{godrich2010target}. 

The modeling choices are justified by the following perturbation error budget for 550 km LEO \cite{vess2003study,klinkrad1998orbit}:
\begin{center}
\footnotesize
\begin{tabular}{lcc}
\hline
\textbf{Perturbation Source} & \textbf{Acceleration (m/s$^2$)} & \textbf{Error at $\Delta t$=1s} \\
\hline
J2 (Earth oblateness) & $\approx 5 \times 10^{-5}$ & 0.025 mm  \\
Atmospheric drag & $10^{-8}$ to $10^{-6}$ & $<$0.001 mm  \\
J3-J6 harmonics & $\approx 10^{-8}$ & $<$0.001 mm  \\
Solar radiation & $\approx 10^{-8}$ & $<$0.001 mm  \\
Third-body (Sun/Moon) & $\approx 5 \times 10^{-7}$ & $<$0.001 mm  \\
\hline
\end{tabular}
\end{center}

\subsection{Unified Impairment Model}

THz transceivers exhibit complex impairments that   deviate from idealized Additive White Gaussian Noise (AWGN) models. We characterize their aggregate effects through three distinct mechanisms \cite{chen2022mcrb}.

\textbf{Phase noise} from frequency multiplication creates integrated variance $\sigma^2_{\phi,\ell}$ that reduces coherent signal power by $e^{-\sigma^2_{\phi,\ell}}$ and introduces power-independent timing jitter:
\begin{equation}
\sigma^2_{\tau,\text{PN}} = \frac{\sigma^2_{\phi,\ell}}{(2\pi f_c)^2}.
\end{equation}

\textbf{Hardware distortions} including PA nonlinearity, I/Q imbalance, and DAC quantization are captured by an effective quality factor $\Gamma_{\text{eff},\ell} = \Gamma_{\text{PA},\ell} + \Gamma_{\text{IQ},\ell} + \Gamma_{\text{DAC},\ell}$. This quality factor directly relates to the standard Error Vector Magnitude (EVM) metric: $\Gamma_{\text{eff}} \approx \text{EVM}^2$ under small distortion conditions, linking our theoretical model to established transceiver specifications \cite{dong2025fundamental}. The resulting signal-dependent distortion power $\sigma^2_{\text{dist},\ell} = \Gamma_{\text{eff},\ell} \cdot P_{\text{signal},\ell}$ creates a normalized distortion term $\text{SNR}_{0,\ell} \cdot \Gamma_{\text{eff},\ell}$ in the SINR denominator,   limiting power scaling benefits.

\textbf{Beam pointing errors} introduce mean power loss $\mathbb{E}[G(\theta)] \approx G_0 \exp(-\sigma^2_{\theta}/\theta^2_{3\text{dB}})$ incorporated into the channel gain, with fast fluctuations contributing to effective noise \cite{dabiri2022pointing,valentini2021analysis}.

These mechanisms synthesize into a unified TOA measurement error variance:
\begin{equation}
\sigma^2_{\text{meas},\ell} = c^2 \left[ \frac{\kappa_{\text{WF}}}{\text{SINR}_{\text{eff},\ell}} + \frac{\sigma^2_{\phi,\ell}}{(2\pi f_c)^2} \right],
\end{equation}
where $\kappa_{\text{WF}} = 1/(8\pi^2\beta^2)$ depends on the RMS bandwidth $\beta$, and the effective SINR incorporating all impairments is:
\begin{equation}
\text{SINR}_{\text{eff},\ell} = \frac{\text{SNR}_{0,\ell} \cdot e^{-\sigma^2_{\phi,\ell}}}{1 + \text{SNR}_{0,\ell} \cdot \Gamma_{\text{eff},\ell} + \sum_{m \neq \ell} \tilde{\alpha}_{\ell m}},
\end{equation}
This model exhibits correct limiting behavior: as $\text{SNR}_{0,\ell} \to \infty$, performance saturates at hardware-determined ceilings, establishing fundamental boundaries unattainable through increased power alone.

\subsection{Aggregated Network Model and Noise Covariance}

The network-wide observation model aggregates all $L$ active ISL measurements:
\begin{equation}
\mathbf{y} = h(\mathbf{x}) + \mathbf{n},
\end{equation}
where $\mathbf{y} = [y_1, \ldots, y_L]^T \in \mathbb{R}^L$, $h(\mathbf{x}) = [h_1(\mathbf{x}), \ldots, h_L(\mathbf{x})]^T$, and $\mathbf{n} = [n_1, \ldots, n_L]^T$.

The noise covariance matrix $\mathbf{C}_n = \mathbb{E}[\mathbf{n}\mathbf{n}^T]$   determines information fusion in the network Fisher Information Matrix. Under independent noise assumptions, $\mathbf{C}_n = \text{diag}(\sigma^2_1, \ldots, \sigma^2_L)$ enables simple FIM decomposition.

However, practical constellations exhibit noise correlations through shared timing references. When satellites derive timing from a common ultra-stable oscillator, the covariance structure becomes:
\begin{equation}
\mathbf{C}_n = \underbrace{\text{diag}(\sigma^2_{1,\text{local}}, \ldots, \sigma^2_{L,\text{local}})}_{\text{Local Noise}} + \sigma^2_{\text{clk}} \mathbf{A}\mathbf{A}^T,
\end{equation}
where the association matrix $\mathbf{A} \in \mathbb{R}^{L \times N_v}$ has rows $[\mathbf{A}]_{\ell,:} = \mathbf{e}_j - \mathbf{e}_i$ for link $\ell$ connecting satellites $i$ and $j$, reflecting the differential nature of clock measurements.

For constellations with shared local oscillator chains or frequency multiplication stages, the covariance structure can be further extended:
\begin{equation}
\mathbf{C}_n = \mathbf{D} + \sigma^2_{\text{clk}} \mathbf{A}\mathbf{A}^T + \sigma^2_{\text{LO}} \mathbf{B}\mathbf{B}^T,
\end{equation}
where $\mathbf{B}$ represents the association matrix for shared phase noise sources, following a similar differential structure as $\mathbf{A}$.

This low-rank structure enables efficient computation via the Woodbury identity:
\begin{equation}
\mathbf{C}_n^{-1} = \mathbf{D}^{-1} - \mathbf{D}^{-1}\mathbf{A}(\mathbf{A}^T\mathbf{D}^{-1}\mathbf{A} + \sigma^{-2}_{\text{clk}}\mathbf{I})^{-1}\mathbf{A}^T\mathbf{D}^{-1},
\end{equation}
requiring only inversion of an $N_v \times N_v$ matrix rather than the full $L \times L$ covariance, providing computational savings when $L \gg N_v$.

The non-diagonal structure embodies fundamental information coupling. In the network FIM $\mathbf{J}_{\text{net}} = \mathbf{H}^T \mathbf{C}_n^{-1} \mathbf{H}$, off-diagonal elements of $\mathbf{C}_n^{-1}$ create cross-terms that couple information from different links. This can enhance performance through common-mode noise cancellation or limit information gain from additional links,   altering the network's scaling behavior with constellation size.

For subsequent performance analysis, the network state can be partitioned as $\mathbf{x} = [\mathbf{x}^T_{\text{kinematic}}, \mathbf{x}^T_{\text{clock}}]^T$, enabling marginalization of clock parameters through the Equivalent Fisher Information Matrix:
\begin{equation}
\mathbf{J}_{\text{EFIM}} = \mathbf{J}_{kk} - \mathbf{J}_{kc} \mathbf{J}_{cc}^{-1} \mathbf{J}_{ck},
\end{equation}
which provides the information content for kinematic states after optimal clock estimation, forming the basis for N-CRLB and GDOP analysis in the following sections.

\section{Network Fisher Information Matrix Framework}

\subsection{Fundamentals of the Information Filter}

Recursive estimation of time-varying LEO-ISAC networks is computationally challenging for traditional Kalman filtering. Dynamic ISL topology changes require intensive matrix restructuring and potentially ill-conditioned inversions, which scale poorly with network size \cite{ramakanth2025centralized}. We adopt the information filter (IF), the dual formulation of the Kalman filter operating in the information domain \cite{kim2003decentralized}.

The IF transforms the estimation problem through the information matrix $\mathbf{J}_k = \mathbf{P}_k^{-1}$ and information vector $\mathbf{y}_k = \mathbf{J}_k\hat{\mathbf{x}}_k$:
\begin{equation}
\mathbf{J}_k = \mathbf{P}_k^{-1}, \quad \mathbf{y}_k = \mathbf{P}_k^{-1}\hat{\mathbf{x}}_k = \mathbf{J}_k\hat{\mathbf{x}}_k,
\label{eq:info_def}
\end{equation}

The fundamental advantage emerges through information additivity. For multiple independent measurements:
\begin{equation}
\mathbf{J}_{\text{total}} = \mathbf{J}_1 + \mathbf{J}_2, \quad \mathbf{y}_{\text{total}} = \mathbf{y}_1 + \mathbf{y}_2,
\label{eq:info_addition}
\end{equation}
contrasting with the covariance update $\mathbf{P}_{\text{total}} = (\mathbf{P}_1^{-1} + \mathbf{P}_2^{-1})^{-1}$ which requires multiple inversions. This additive property is well-suited for dynamic networks: when ISL $\ell$ becomes active, its contribution seamlessly integrates as:
\begin{equation}
\mathbf{J}_{\text{net}}(k|k) = \mathbf{J}_{\text{net}}(k|k-1) + \mathbf{J}_{\ell}
\label{eq:link_addition}.
\end{equation}
No matrix restructuring is needed when links deactivate—their absence simply excludes their contribution. The IF also effectively handles initialization with no prior knowledge through $\mathbf{J}_0 = \mathbf{0}$, avoiding numerical complications of infinite covariance representation.
\subsection{Time Update (Prediction)}

This prediction step transforms posterior information at time $k-1$ into prior information at time $k$, utilizing the linearized state transition model:
\begin{equation}
\mathbf{x}_k = \mathbf{F}_k \mathbf{x}_{k-1} + \mathbf{w}_{k-1}, \quad \mathbf{w}_{k-1} \sim \mathcal{N}(0, \mathbf{Q}_{k-1}).
\label{eq:state_transition}
\end{equation}

To ensure numerical stability and avoid inversion of potentially singular transition matrices, we employ a formulation based on the Woodbury matrix identity:
\begin{align}
\mathbf{M}_k &= \left( \mathbf{J}_{\text{net}}(k-1|k-1) + \mathbf{F}_k^T \mathbf{Q}_{k-1}^{-1} \mathbf{F}_k \right)^{-1} \label{eq:M_matrix},\\
\mathbf{J}_{\text{net}}(k|k-1) &= \mathbf{Q}_{k-1}^{-1} - \mathbf{Q}_{k-1}^{-1} \mathbf{F}_k \mathbf{M}_k \mathbf{F}_k^T \mathbf{Q}_{k-1}^{-1} \label{eq:J_predict},\\
\mathbf{y}_{\text{net}}(k|k-1) &= \mathbf{J}_{\text{net}}(k|k-1) \mathbf{F}_k \mathbf{J}_{\text{net}}^{-1}(k-1|k-1) \mathbf{y}_{\text{net}}(k-1|k-1) .\label{eq:y_predict}
\end{align}

\noindent\textit{Note: In practical implementation, use linear equation solving $\hat{\mathbf{x}} = \text{solve}(\mathbf{J}, \mathbf{y})$ rather than explicit matrix inversion $\hat{\mathbf{x}} = \mathbf{J}^{-1}\mathbf{y}$ for improved numerical stability.}

The structure of \eqref{eq:J_predict} reveals how temporal information propagates. The first term $\mathbf{Q}_{k-1}^{-1}$ represents the maximum information that could be retained with perfect state transition knowledge. The Woodbury correction term quantifies information loss due to propagation uncertainty. This manifests as "information depreciation"—process noise injection increases differential entropy, corresponding to reduced Fisher information content. Under typical conditions with process noise present, we observe $\text{tr}(\mathbf{J}_{\text{net}}(k|k-1)) < \text{tr}(\mathbf{J}_{\text{net}}(k-1|k-1))$, indicating net information loss \cite{masri2016recursive}. 

For our block-diagonal state transition structure where $\mathbf{F}_k = \text{diag}(\mathbf{F}_{1,k}, \ldots, \mathbf{F}_{N_v,k})$, the prediction naturally decomposes into parallel operations for each satellite, providing significant computational advantages for large constellations \cite{olfati2007consensus}.

\subsection{Measurement Update (Correction)}

Following temporal propagation, the measurement update incorporates all active ISL observations to refine the state estimate. This correction step exemplifies the information filter's utility: fusion of multiple measurements reduces to elementary algebraic operations:
\begin{align}
\mathbf{J}_{\text{net}}(k|k) &= \mathbf{J}_{\text{net}}(k|k-1) + \sum_{\ell \in \mathcal{E}(t_k)} \mathbf{J}_\ell \label{eq:J_update},\\
\mathbf{y}_{\text{net}}(k|k) &= \mathbf{y}_{\text{net}}(k|k-1) + \sum_{\ell \in \mathcal{E}(t_k)} \mathbf{H}_\ell^T \mathbf{R}_\ell^{-1} \mathbf{z}_{\ell,k} .\label{eq:y_update}
\end{align}

Each active ISL $\ell$ contributes an information matrix increment:
\begin{equation}
\mathbf{J}_\ell = \mathbf{H}_\ell^T \mathbf{R}_\ell^{-1} \mathbf{H}_\ell
\label{eq:link_fim},
\end{equation}
where $\mathbf{H}_\ell$ is the sparse measurement Jacobian from Section 2.2, and $\mathbf{R}_\ell = \sigma^2_{\text{meas},\ell}$ represents the measurement noise variance.

Consistent with the unified physical model established in Section 2.3, the measurement error variance for TOA-based ranging incorporates both hardware-limited and phase-noise-limited regimes:
\begin{equation}
\sigma^2_{\text{meas},\ell} = c^2 \left[ \frac{\kappa_{\text{WF}}}{\text{SINR}_{\text{eff},\ell}} + \frac{\sigma^2_{\phi,\ell}}{(2\pi f_c)^2} \right]
\label{eq:noise_model_unified},
\end{equation}
where $\kappa_{\text{WF}} = 1/(8\pi^2\beta^2)$ is the waveform-dependent constant with $\beta$ denoting the effective RMS bandwidth. The pre-impairment SNR is defined as $\text{SNR}_{0,\ell} = P_\ell|g_\ell|^2/N_0$, and the effective SINR from Section 2.3 is:
\begin{equation}
\text{SINR}_{\text{eff},\ell} = \frac{\text{SNR}_{0,\ell} \cdot e^{-\sigma^2_{\phi,\ell}}}{1 + \text{SNR}_{0,\ell} \cdot \Gamma_{\text{eff},\ell} + \sum_{m \neq \ell} \tilde{\alpha}_{\ell m}}.
\label{eq:sinr_eff_recall}
\end{equation}
This formulation ensures consistency between our sensing performance model (CRLB) and communication performance model (capacity), both built upon the same unified physical impairment framework.

The summation formulation in \eqref{eq:J_update}-\eqref{eq:y_update} assumes independent measurement noise across links. When noise correlations exist—such as from shared timing references discussed in Section 2.4—the measurement update requires a block form:
\begin{equation}
\mathbf{J}_{\text{net}}(k|k) = \mathbf{J}_{\text{net}}(k|k-1) + \mathbf{H}^T \mathbf{C}_n^{-1} \mathbf{H}
\label{eq:J_update_correlated},
\end{equation}
where $\mathbf{H}$ stacks all measurement Jacobians and $\mathbf{C}_n$ is the full noise covariance matrix including off-diagonal correlations. As shown in Section 2.4, $\mathbf{C}_n^{-1}$ can be efficiently computed via the Woodbury identity when correlations arise from low-rank structures like shared clocks.

The summation formulation provides an inherently effective solution to dynamic topology challenges. When satellites enter mutual visibility, their information contribution simply appears as a new term. When links become inactive, their absence naturally excludes their contribution—no complex bookkeeping required. This represents "information injection" where each ISL deposits knowledge proportional to its measurement quality ($\mathbf{R}_\ell^{-1}$) and geometric leverage ($\mathbf{H}_\ell$) \cite{shen2010fundamental}.

For hardware-limited THz systems, the information contribution saturates at high power. Taking the limit of \eqref{eq:noise_model_unified} as $\text{SNR}_{0,\ell} \to \infty$:
\begin{equation}
\mathbf{J}_{\ell,\text{sat}} = \frac{\mathbf{H}_\ell^T \mathbf{H}_\ell}{c^2 \left[ \kappa_{\text{WF}} \cdot \Gamma_{\text{eff},\ell} \cdot e^{\sigma^2_{\phi,\ell}} + \frac{\sigma^2_{\phi,\ell}}{(2\pi f_c)^2} \right]}
\label{eq:J_saturate}.
\end{equation}
This saturation value directly emerges from the limiting behavior of our unified $\sigma^2_{\text{meas},\ell}$ model,   bounding extractable information regardless of power allocation. The update maintains monotonicity ($\mathbf{J}_{\text{net}}(k|k) \succeq \mathbf{J}_{\text{net}}(k|k-1)$), ensuring measurements only increase information content .

\subsection{Parameter Marginalization via the Equivalent FIM (EFIM)}

Practical sensing applications require marginalizing nuisance clock parameters to focus on kinematic states. Partitioning the state vector as $\mathbf{x} = [\mathbf{a}^T, \mathbf{b}^T]^T$ with kinematic states $\mathbf{a} \in \mathbb{R}^{6N_v}$ and clock parameters $\mathbf{b} \in \mathbb{R}^{2N_v}$, the network FIM becomes:
\begin{equation}
\mathbf{J}_{\text{net}} = \begin{bmatrix} 
\mathbf{J}_{\mathbf{aa}} & \mathbf{J}_{\mathbf{ab}} \\ 
\mathbf{J}_{\mathbf{ba}} & \mathbf{J}_{\mathbf{bb}} 
\end{bmatrix}.
\label{eq:partition}
\end{equation}

The Equivalent Fisher Information Matrix obtained via the Schur complement:
\begin{equation}
\mathbf{J}_{\text{EFIM}}(\mathbf{a}) = \mathbf{J}_{\mathbf{aa}} - \mathbf{J}_{\mathbf{ab}} \mathbf{J}_{\mathbf{bb}}^{-1} \mathbf{J}_{\mathbf{ba}}
\label{eq:efim},
\end{equation}
reveals fundamental physical insight: $\mathbf{J}_{\mathbf{aa}}$ represents the idealized information with perfect clocks (unattainable upper bound), while the subtracted term quantifies information penalty from clock uncertainty propagating through parameter coupling.

Network cooperation mitigates this penalty through collective clock observability:
\begin{equation}
\mathbf{J}_{\mathbf{bb}} = \sum_{\ell \in \mathcal{E}(t_k)} \mathbf{J}_{\mathbf{bb}}^{(\ell)}
\label{eq:clock_info}.
\end{equation}
As connectivity increases, $\mathbf{J}_{\mathbf{bb}}$ grows linearly, reducing $\mathbf{J}_{\mathbf{bb}}^{-1}$ and embodying "cooperation as calibration." This physical mechanism emerges from the structure of clock observability: each ISL measurement $h_\ell = \frac{1}{c}\|\mathbf{p}_j - \mathbf{p}_i\| + (b_j - b_i)$ observes only clock differences, contributing a rank-deficient matrix $\mathbf{J}_{\mathbf{bb}}^{(\ell)}$ to the aggregate. However, summing these contributions over the network graph:
\begin{equation}
\mathbf{J}_{\mathbf{bb}} = \sum_{\ell \in \mathcal{E}} \mathbf{J}_{\mathbf{bb}}^{(\ell)} = \mathbf{L}_G \otimes \sigma^{-2}_{\text{clock}},
\end{equation}
where $\mathbf{L}_G$ is the graph Laplacian, yields a matrix with rank $N_v-1$—full observability except for the global time offset. This mathematical aggregation enables the network to resolve all relative clock states through geometric diversity, minimizing the information penalty $\mathbf{J}_{\mathbf{ab}} \mathbf{J}_{\mathbf{bb}}^{-1} \mathbf{J}_{\mathbf{ba}}$ in the EFIM. The Network Cramér-Rao Lower Bound follows as $\text{CRLB}_{\text{net}}(\mathbf{a}) = \mathbf{J}_{\text{EFIM}}(\mathbf{a})^{-1}$, with spectral properties providing direct observability metrics.

\section{Network Interference and Opportunistic Sensing}

\subsection{A Geometric-Stochastic Interference Model}

Networked THz LEO-ISAC systems face co-channel interference that   differs from additive white Gaussian noise. The interference exhibits structured patterns determined by deterministic network geometry and stochastic platform dynamics—particularly pointing jitter that creates intermittent high-power coupling events. This section develops the physical model capturing these effects.

\subsubsection{Gaussian Beam Approximation for THz Antennas}

At THz frequencies, satellite antennas achieve high gain through narrow beamwidths on the order of millidegrees. The radiation pattern is well-approximated by the fundamental Gaussian beam mode (TEM$_{00}$) \cite{alda2003laser}, with intensity distribution:
\begin{equation}
I(r,z) = \frac{2P_t}{\pi w(z)^2} \exp\left(-\frac{2r^2}{w(z)^2}\right)
\label{eq:gaussian_intensity},
\end{equation}
where $P_t$ is transmitted power and $w(z)$ is the beam radius at distance $z$.

The beam radius evolution follows:
\begin{equation}
w(z) = w_0 \sqrt{1 + \left(\frac{\lambda z}{\pi w_0^2}\right)^2}
\label{eq:beam_evolution}.
\end{equation}
For ISLs in the far-field regime ($z \gg \pi w_0^2/\lambda$), this simplifies to $w(z) \approx \lambda z/(\pi w_0)$. The half-power beamwidth relates to beam parameters as:
\begin{equation}
\theta_B \approx \frac{\lambda}{\pi w_0}\sqrt{2\ln 2}
\label{eq:beamwidth}.
\end{equation}
This relationship connects antenna design parameters to the spatial interference pattern.

\subsubsection{Pointing Error Characterization}

The narrow beamwidths enabling frequency reuse also introduce high sensitivity to pointing errors from platform vibrations and tracking imperfections \cite{he2022analysis}. These deviations transform negligible sidelobe leakage into potentially strong interference events.

Pointing errors in azimuth and elevation are modeled as independent zero-mean Gaussian variables:
\begin{align}
\theta_{\text{az}} &\sim \mathcal{N}(0, \sigma_e^2), \\
\theta_{\text{el}} &\sim \mathcal{N}(0, \sigma_e^2),
\label{eq:pointing_components}
\end{align}
where $\sigma_e^2$ represents the pointing error variance. The framework extends to anisotropic jitter by replacing $2\sigma_e^2$ with the appropriate second-order moment.

The radial pointing error:
\begin{equation}
\theta_e = \sqrt{\theta_{\text{az}}^2 + \theta_{\text{el}}^2},
\label{eq:radial_error}
\end{equation}
follows a Rayleigh distribution as the norm of two independent Gaussians \cite{valentini2021analysis}:
\begin{equation}
f_{\theta_e}(\theta) = \frac{\theta}{\sigma_e^2} \exp\left(-\frac{\theta^2}{2\sigma_e^2}\right), \quad \theta \geq 0,
\label{eq:rayleigh_pdf}
\end{equation}
with mean $\mathbb{E}[\theta_e] = \sigma_e\sqrt{\pi/2}$ and second moment $\mathbb{E}[\theta_e^2] = 2\sigma_e^2$, directly linking platform stability to pointing performance.

\subsubsection{Average Received Interference Power}

The instantaneous interference power received at victim receiver $\ell$ from interfering transmitter $m$ depends on both satellites' pointing errors. For a circular receiving aperture of radius $a_r$ located at distance $d_{\ell m}$, the instantaneous received power is:
\begin{equation}
P_r(\theta_{e,m}, \theta_{e,\ell}) = \iint_{A_r} I(r', d_{\ell m}) \, dA,
\label{eq:instant_power}
\end{equation}
where $r'$ is the distance from the integration point to the actual beam center, which itself is randomly displaced due to pointing errors.

The average received interference power requires integration over all possible pointing error realizations:
\begin{equation}
\langle P_r \rangle_{\ell m} = \int_0^{2\pi} \int_0^{\infty} P_r(\theta_e, \phi_e) \cdot \frac{\theta_e}{2\pi\sigma_e^2} \exp\left(-\frac{\theta_e^2}{2\sigma_e^2}\right) \, d\theta_e \, d\phi_e.
\label{eq:average_power}
\end{equation}

This four-fold integral—two from the aperture integration and two from the statistical averaging—captures the complex interplay between deterministic beam propagation and stochastic pointing dynamics. While direct analytical evaluation is intractable, the Gaussian nature of both the beam profile and pointing statistics enables powerful approximation techniques.

Under the point receiver approximation valid when $a_r \ll w(d_{\ell m})$, the received power becomes proportional to the intensity at the receiver location:
\begin{equation}
P_r \approx A_{\text{eff}} \cdot I(d_{\ell m}\theta_{\text{dev}}, d_{\ell m}),
\label{eq:point_receiver}
\end{equation}
where $A_{\text{eff}} = G_R\lambda^2/(4\pi)$ is the effective aperture area and $\theta_{\text{dev}}$ is the total angular deviation between transmitter boresight and receiver location.

This geometric-stochastic framework reveals a fundamental characteristic of THz ISL interference. It is dominated not by steady-state sidelobe coupling, but by transient "glint" events where pointing errors cause main beams to momentarily sweep across unintended receivers \cite{niu2024interference}. The Rayleigh-distributed pointing errors transform the interference from a constant background into a highly dynamic, burst-like process. 

The statistical nature of these glint events follows from the tail distribution of the total angular deviation, which exhibits non-central chi-squared characteristics. While exact closed-form expressions for exceedance probabilities are complex, the behavior is   governed by exponential decay with characteristic angular scale $\sqrt{\theta_B^2/2 + 2\sigma_e^2}$—the same effective beam broadening appearing in our interference coefficient. This burst-like interference has implications for protocol design, as traditional interference mitigation strategies assuming steady-state noise become ineffective against these intermittent high-power events.

\subsection{Closed-Form Approximation of the Interference Coefficient}

The average received interference power requires integrating over both aperture and pointing error distributions. For THz ISLs where beam radius far exceeds antenna apertures ($a_r \ll w(d_{\ell m})$), the point-receiver approximation applies \cite{carrillo2023effects}.

The instantaneous received power at receiver $\ell$ from interferer $m$ becomes:
\begin{equation}
P_r(\theta_e, \phi_e) \approx A_{\text{eff}} \cdot \frac{2P_m}{\pi w(d_{\ell m})^2} \exp\left(-\frac{2d_{\ell m}^2 \theta_{\text{dev}}^2}{w(d_{\ell m})^2}\right),
\label{eq:instant_power_approx}
\end{equation}
where $A_{\text{eff}} = G_{R,\ell}\lambda^2/(4\pi)$ is the effective aperture and the total angular deviation is:
\begin{equation}
\theta_{\text{dev}}^2 = \theta_{\ell m}^2 + \theta_e^2 - 2\theta_{\ell m}\theta_e\cos\phi_e
\label{eq:angular_deviation},
\end{equation}
with $\theta_{\ell m}$ being the geometric misalignment between nominal pointing directions.

Averaging over the Rayleigh-distributed pointing error yields:
\begin{equation}
\langle P_r \rangle_{\ell m} = \int_0^{2\pi} \int_0^{\infty} P_r(\theta, \phi) \cdot \frac{\theta}{2\pi\sigma_e^2} \exp\left(-\frac{\theta^2}{2\sigma_e^2}\right) \, d\theta \, d\phi
\label{eq:statistical_average}.
\end{equation}

The integral evaluates analytically via Bessel function techniques, yielding:
\begin{equation}
\begin{aligned}
\langle P_r \rangle_{\ell m} &= P_m G_{T,m} G_{R,\ell} \left(\frac{\lambda}{4\pi d_{\ell m}}\right)^2 \\
&\quad \cdot \frac{1}{1 + 4\sigma_{e,m}^2/\theta_{B,m}^2} \cdot \exp\left(-\frac{\theta_{\ell m}^2}{\theta_{B,m}^2/2 + 2\sigma_{e,m}^2}\right).
\end{aligned}
\label{eq:avg_power_closed_general}
\end{equation}

The interference coefficient $\alpha_{\ell m} = \langle P_r \rangle_{\ell m}/\langle P_s \rangle_\ell$ quantifies the interference-to-signal ratio. For homogeneous constellations with identical hardware:
\begin{equation}
\alpha_{\ell m} = \left(\frac{d_\ell}{d_{\ell m}}\right)^2 \exp\left(-\frac{\theta_{\ell m}^2}{\theta_B^2/2 + 2\sigma_e^2}\right).
\label{eq:interference_coefficient_simple}
\end{equation}

The closed-form approximation in \eqref{eq:interference_coefficient_simple} is highly accurate under typical ISL conditions, namely the far-field, small-angle ($\theta_{\ell m} \lesssim 3\theta_B$), and point-receiver ($a_r \ll w(d_{\ell m})$) regimes. A Monte Carlo validation comparing this analytical model to numerical integration of \eqref{eq:average_power} shows a relative error of less than 5\% within this domain, confirming its suitability for large-scale network analysis.

The denominator $\theta_B^2/2 + 2\sigma_e^2$ in the exponential reveals a fundamental physical insight: pointing jitter effectively broadens the antenna beam pattern. The term $\theta_B^2/2$ represents the inherent beam divergence, while $2\sigma_e^2$ captures the additional angular spreading due to platform instability \cite{song2017impact}.

This effective broadening has significant implications for system design:
\begin{itemize}
\item \textbf{Static Regime} ($\sigma_e \ll \theta_B$): When platform stability is excellent relative to beamwidth, interference is dominated by the antenna pattern itself. The interference coefficient follows a Gaussian decay with characteristic width $\theta_B/\sqrt{2}$.

\item \textbf{Dynamic Regime} ($\sigma_e \gg \theta_B$): When pointing jitter dominates, the effective beam becomes much wider than designed. Interference extends to larger angular separations, but with reduced peak intensity due to power spreading.

\item \textbf{Design Trade-off}: Narrowing the beam ($\theta_B \downarrow$) improves interference isolation when pointing is stable but increases sensitivity to jitter. This quantitative relationship enables joint optimization of antenna design and platform specifications.
\end{itemize}

Incorporating the derived interference coefficient into the effective SINR framework established in Section II, we obtain the complete network performance model consistent with the noise-normalized formulation used throughout this work:
\begin{equation}
\text{SINR}_{\text{eff},\ell} = \frac{\text{SNR}_{0,\ell} \cdot e^{-\sigma^2_{\phi,\ell}}}{1 + \text{SNR}_{0,\ell} \cdot \Gamma_{\text{eff},\ell} + \sum_{m \neq \ell} \tilde{\alpha}_{\ell m}}
\label{eq:network_sinr_unified},
\end{equation}
where the normalized interference coefficient $\tilde{\alpha}_{\ell m}$ is defined as:
\begin{equation}
\tilde{\alpha}_{\ell m} \triangleq \text{SNR}_{0,\ell} \cdot \alpha_{\ell m}
\label{eq:normalized_interference},
\end{equation}
with $\alpha_{\ell m}$ being the interference-to-signal power ratio given by \eqref{eq:interference_coefficient_simple} for homogeneous networks. This ensures consistency with the noise-normalized denominator structure where all terms represent power ratios relative to the thermal noise floor $N_0$.

The network interference term is thus:
\begin{equation}
\sum_{m \neq \ell} \tilde{\alpha}_{\ell m} = \text{SNR}_{0,\ell} \sum_{m \neq \ell} \alpha_{\ell m}
\label{eq:network_interference_normalized}.
\end{equation}

This unified expression enables rapid evaluation of interference levels across the entire constellation. The model reveals three distinct operating regimes based on the dominant term in the SINR denominator:
\begin{enumerate}
\item \textbf{Noise-limited}: When $1 \gg \text{SNR}_{0,\ell} \cdot \Gamma_{\text{eff},\ell} + \sum_{m \neq \ell} \tilde{\alpha}_{\ell m}$, thermal noise dominates.

\item \textbf{Hardware-limited}: When $\text{SNR}_{0,\ell} \cdot \Gamma_{\text{eff},\ell} \gg 1 + \sum_{m \neq \ell} \tilde{\alpha}_{\ell m}$, self-interference from hardware impairments becomes the bottleneck.

\item \textbf{Interference-limited}: When $\sum_{m \neq \ell} \tilde{\alpha}_{\ell m} \gg 1 + \text{SNR}_{0,\ell} \cdot \Gamma_{\text{eff},\ell}$, network interference dominates.
\end{enumerate}

This unified model provides both the analytical foundation for performance evaluation and clear guidance for system optimization based on the identified limiting factor.

\subsection{Quantifying Information Gain from Opportunistic Sensing}

The ISAC paradigm   transforms our perspective on interference: rather than viewing co-channel signals as purely detrimental noise to be suppressed, they become potential sources of valuable sensing information. When an interfering signal illuminates a target and its scattered echo reaches an unintended receiver, this creates an opportunistic bistatic radar configuration. This section quantifies the sensing information gain from such opportunistic measurements through the rigorous framework of Fisher Information theory.

\subsubsection{Bistatic Radar Model for Opportunistic Links}

Consider an opportunistic sensing link where transmitter $m$ (originally causing interference to receiver $\ell$) inadvertently illuminates a target at position $\mathbf{p}_t$. The scattered signal received at $\ell$ forms a bistatic radar system with the following geometry \cite{wang2022integrated}:

\begin{itemize}
\item Transmitter: Satellite $m$ at position $\mathbf{p}_m$
\item Receiver: Satellite $\ell$ at position $\mathbf{p}_\ell$
\item Target: Object at unknown position $\mathbf{p}_t = [x_t, y_t, z_t]^T$
\end{itemize}

The fundamental measurement in this bistatic configuration is the bistatic range, defined as the total signal propagation path:
\begin{equation}
R_b(\mathbf{p}_t) = \|\mathbf{p}_m - \mathbf{p}_t\| + \|\mathbf{p}_t - \mathbf{p}_\ell\|
\label{eq:bistatic_range}.
\end{equation}

This measurement constrains the target to lie on an ellipsoid with foci at $\mathbf{p}_m$ and $\mathbf{p}_\ell$. The precision of this constraint depends on both the signal-to-noise ratio of the echo and the geometric configuration.

\subsubsection{Fisher Information Matrix Framework}

The Fisher Information Matrix (FIM) provides the fundamental tool for quantifying the information content of measurements about unknown parameters. For an unbiased estimator of parameter vector $\boldsymbol{\theta}$, the Cramér-Rao Lower Bound (CRLB) theorem establishes that the estimation error covariance satisfies:
\begin{equation}
\text{Cov}(\hat{\boldsymbol{\theta}}) \geq \mathbf{J}^{-1}(\boldsymbol{\theta})
\label{eq:crlb_inequality}.
\end{equation}
where $\mathbf{J}(\boldsymbol{\theta})$ is the FIM. Thus, larger FIM values indicate more information and potentially better estimation performance.

\subsubsection{Derivation of Opportunistic Sensing FIM}

The FIM for bistatic range measurement is expressed in its standard variance-driven form, maintaining consistency with the unified ISAC model developed in Section II:
\begin{equation}
\mathbf{J}_{\text{IoO}}(\mathbf{p}_t) = \frac{1}{\sigma_{R_b}^2} \left(\nabla_{\mathbf{p}_t} R_b\right) \left(\nabla_{\mathbf{p}_t} R_b\right)^T
\label{eq:ioo_fim_standard},
\end{equation}
where the bistatic ranging error variance $\sigma_{R_b}^2$ follows the same physical structure as the direct-path TOA measurement variance:
\begin{equation}
\sigma_{R_b}^2 = c^2 \left[ \frac{\kappa_{\text{WF}}}{\text{SINR}_{\text{IoO}}} + \frac{\sigma^2_{\phi}}{(2\pi f_c)^2} \right]
\label{eq:bistatic_variance}.
\end{equation}

Here, $\text{SINR}_{\text{IoO}}$ is the effective signal-to-interference-plus-noise ratio of the echo signal, given by the bistatic radar equation:
\begin{equation}
\text{SINR}_{\text{IoO}} = \frac{P_t G_{T,m}(\theta_{m,t}) G_{R,\ell}(\theta_{\ell,t}) \sigma_b \lambda^2}{(4\pi)^3 R_{m,t}^2 R_{\ell,t}^2 L_{\text{proc}} N_{\text{eff}}}
\label{eq:bistatic_sinr}
\end{equation}
where $\sigma_b$ is the bistatic radar cross-section, $G_{T,m}(\theta_{m,t})$ and $G_{R,\ell}(\theta_{\ell,t})$ are antenna gains in the target direction, $L_{\text{proc}}$ represents processing losses, and $N_{\text{eff}}$ is the effective noise power including residual direct-path interference.

The gradient of bistatic range with respect to target position is:
\begin{align}
\nabla_{\mathbf{p}_t} R_b &= \frac{\mathbf{p}_t - \mathbf{p}_m}{\|\mathbf{p}_t - \mathbf{p}_m\|} + \frac{\mathbf{p}_t - \mathbf{p}_\ell}{\|\mathbf{p}_t - \mathbf{p}_\ell\|} \nonumber \\
&= \mathbf{u}_{m \to t} + \mathbf{u}_{\ell \to t}
\label{eq:bistatic_gradient},
\end{align}
where $\mathbf{u}_{m \to t}$ and $\mathbf{u}_{\ell \to t}$ are unit vectors pointing from the transmitter and receiver to the target, respectively.

The geometric matrix becomes \cite{temiz2022range}:
\begin{equation}
\mathbf{G}_{\text{geom}} = (\nabla_{\mathbf{p}_t} R_b)(\nabla_{\mathbf{p}_t} R_b)^T = (\mathbf{u}_{m \to t} + \mathbf{u}_{\ell \to t})(\mathbf{u}_{m \to t} + \mathbf{u}_{\ell \to t})^T,
\label{eq:geometric_matrix}.
\end{equation}

This $3 \times 3$ matrix has rank at most 1, with eigenvalue $\lambda_1 = 2(1 + \cos\beta)$ where $\beta$ is the bistatic angle, indicating that a single bistatic measurement provides information only along the bistatic bisector direction.

\subsubsection{Information Fusion and Network Enhancement}

The fundamental principle of information fusion for independent measurements is the additivity of Fisher Information Matrices. When the opportunistic sensing capability is activated, the network's total FIM for target state estimation updates according to:
\begin{equation}
\mathbf{J}_{\text{net,new}} = \mathbf{J}_{\text{net,old}} + \mathbf{J}_{\text{IoO}}
\label{eq:fim_fusion}.
\end{equation}

This deceptively simple addition rule reveals the significant potential of networked ISAC systems:
\textbf{1. Geometric Diversity:} While a single bistatic measurement provides information only along one direction (rank-1 FIM), combining multiple opportunistic links with different geometric configurations creates a full-rank FIM.

\textbf{2. Weak Signal Aggregation:} Even opportunistic links with low SNR contribute positively to the network FIM. Many weak measurements can collectively provide substantial information.

\textbf{3. Robustness Through Redundancy:} The reduction in the volume of the error ellipsoid is governed by the matrix determinant lemma for rank-1 updates. The ratio of the new volume to the old volume is:
\begin{equation}
\frac{\text{Vol}_{\text{new}}}{\text{Vol}_{\text{old}}} = \frac{1}{\sqrt{1 + \frac{1}{\sigma_{R_b}^2} (\nabla_{\mathbf{p}_t} R_b)^T \mathbf{J}_{\text{net,old}}^{-1} (\nabla_{\mathbf{p}_t} R_b)}}
\label{eq:volume_reduction_correct}.
\end{equation}

This equation reveals a  key insight: the information gain is maximized when the new measurement's gradient vector $\nabla_{\mathbf{p}_t} R_b$ is aligned with the direction of maximum uncertainty of the existing network, i.e., the principal eigenvector of the prior CRLB, $\mathbf{J}_{\text{net,old}}^{-1}$.

\subsubsection{Practical Considerations}

While the information-theoretic analysis reveals substantial gains, practical implementation faces challenges. Signal separability requires sophisticated processing analogous to full-duplex self-interference cancellation to handle the 40-60 dB dynamic range between direct and scattered paths. Precise nanosecond-level time synchronization between non-cooperating nodes is essential for meter-level ranging precision. Statistical correlation between the opportunistic echo and direct-path interference, originating from the same source, may necessitate robust fusion techniques like Covariance Intersection (CI) to ensure consistent state estimation. Additionally, matched filtering requires knowledge or estimation of transmitted waveform parameters through either explicit coordination or blind estimation techniques.

Despite these challenges, the quantified information gains—demonstrated as 5.67 dB improvement under specific conditions—justify pursuing opportunistic sensing as a key capability. The transformation of interference from performance limiter to information source represents a fundamental paradigm shift enabled by our analytical framework.

\section{Numerical Validation and Performance Analysis}
This section validates our theoretical framework through comprehensive numerical evaluation at representative static configurations. By analyzing instantaneous N-CRLB using the measurement update and EFIM marginalization components of our recursive framework, we isolate fundamental trade-offs between system parameters. While full orbital-period dynamic simulation represents a natural extension, this foundational analysis establishes the key performance boundaries and design principles governing THz LEO-ISAC networks.

\subsection{Simulation Setup}

We validate our theoretical framework through   numerical simulations using realistic LEO constellation parameters. Table~\ref{tab:sim_params} summarizes the key simulation parameters.

\begin{table}[h]
\centering
\caption{Simulation Parameters}
\label{tab:sim_params}
\begin{tabular}{llc}
\hline
\textbf{Category} & \textbf{Parameter} & \textbf{Value} \\
\hline
\multirow{4}{*}{Constellation} 
& Number of satellites & 4--8 \\
& Orbital altitude & 550 km \\
& ISL range limit & 8,000 km \\
& Update interval & 1 s \\
\hline
\multirow{5}{*}{THz Signal} 
& Carrier frequency $f_c$ & 300 GHz \\
& Bandwidth $B$ & 10 GHz \\
& Transmit power $P_t$ & 10 W \\
& Antenna gain $G_T, G_R$ & 50 dBi \\
& Beamwidth (HPBW) $\theta_B$ & 14 mdeg \\
\hline
\multirow{3}{*}{Pointing} 
& Pointing error $\sigma_e$ & 2.8 mdeg \\
& Jitter bandwidth & 100 Hz \\
& Stability ratio $\sigma_e/\theta_B$ & 0.2 \\
\hline
\multirow{3}{*}{Clock} 
& Clock bias drift & $10^{-20}$ s$^2$/s \\
& Clock rate drift & $10^{-26}$ s$^2$/s$^3$ \\
& Correlation coefficient $\rho$ & 0--0.96 \\
\hline
\end{tabular}
\end{table}

Four hardware profiles capture different technology maturity levels, characterized by the effective quality factor $\Gamma_{\text{eff}}$ and phase noise variance $\sigma^2_\phi$:

\begin{table*}[h]
\centering
\caption{Hardware Impairment Profiles mapping to quality factor $\Gamma_{\text{eff}}$, phase noise $\sigma^2_\phi$, and noise figure (NF)}
\label{tab:hardware}
\begin{tabular}{lcccc}
\hline
\textbf{Profile} & \textbf{State-of-the-Art} & \textbf{High-Performance} & \textbf{SWaP-Efficient} & \textbf{Low-Cost} \\
\hline
$\Gamma_{\text{eff}}$ & 0.005 & 0.01 & 0.045 & 0.05 \\
$\sigma^2_\phi$ & $10^{-4}$ & $10^{-3}$ & $10^{-2}$ & $10^{-1}$ \\
Noise figure (dB) & 5 & 7 & 10 & 12 \\
Technology & InP & InP/SiGe & SiGe & Silicon \\
\hline
\end{tabular}
\end{table*}

Three constellation geometries are evaluated: (i) \textit{Planar}—satellites uniformly distributed in a single orbital plane, (ii) \textit{Cubic}—satellites at cube vertices for 3D diversity, and (iii) \textit{Random}—stochastic perturbations with $\sigma_r = 100$ km modeling orbital uncertainties. For opportunistic sensing analysis, we specifically simulate a geometrically challenged scenario where three of four satellites are nearly coplanar, creating severe vertical observability deficiency with error ellipsoid elongation exceeding 10:1.

The effective SINR incorporating all impairments follows ~\eqref{eq:sinr_eff_recall}, with normalized interference $\tilde{\alpha} = \sum_{m \neq \ell} \alpha_{\ell m}$ ranging from $10^{-3}$ (sparse) to $10^{0}$ (dense) networks. Monte Carlo simulations use 1,000 realizations with fixed random seeds for reproducibility. All implementations use Python 3.10 with NumPy/SciPy, employing the Woodbury identity for efficient matrix updates and regularization $\epsilon = 10^{-10}$ for numerical stability.

\subsection{Fundamental Physical Limits}

\begin{figure}[!t]
    \centering
    \includegraphics[width=0.9\columnwidth]{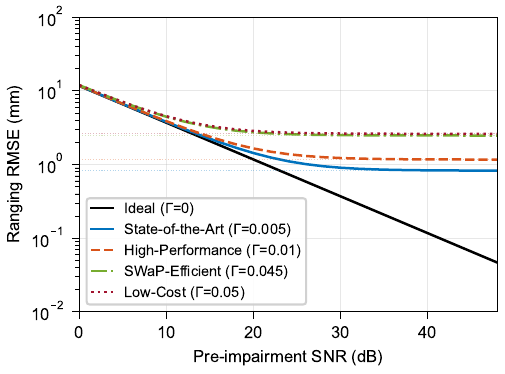}
\caption{Ranging RMSE vs. SNR for different hardware quality levels ($\Gamma_{\text{eff}}$), showing hardware-limited performance ceilings.}

    \label{fig:hw_ceiling}
\end{figure}

Fig.~\ref{fig:hw_ceiling} validates the hardware-imposed performance ceiling. The effective SINR saturates at $e^{-\sigma^2_\phi}/\Gamma_{\text{eff}}$ for $\text{SNR}_0 \gg 1/\Gamma_{\text{eff}}$, creating horizontal asymptotes: 0.83 mm for state-of-the-art hardware ($\Gamma_{\text{eff}} = 0.005$) versus 2.61 mm for low-cost implementations ($\Gamma_{\text{eff}} = 0.05$). The critical transition at $\text{SNR}_{\text{crit}} = 1/\Gamma_{\text{eff}}$   limits power-scaling benefits.

\begin{figure}[!t]
    \centering
    \includegraphics[width=0.9\columnwidth]{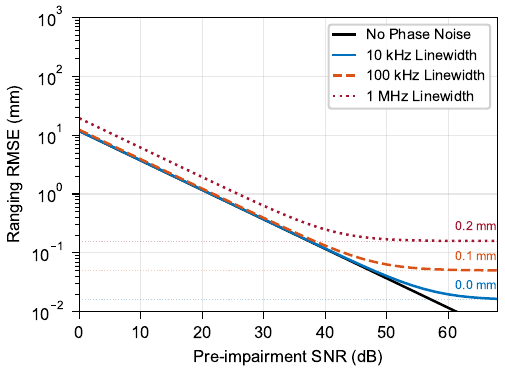}
\caption{Ranging RMSE vs. SNR for different phase noise levels, showing power-independent error floors.}

    \label{fig:pn_floor}
\end{figure}

Fig.~\ref{fig:pn_floor} confirms the power-independent error floor $\sigma_{\text{floor}} = c\sqrt{\sigma^2_\phi}/(2\pi f_c)$. At 300 GHz, measured floors match theory: 0.051 mm (100 kHz linewidth) and 0.16 mm (1 MHz linewidth). The achievable precision bound $\sigma_{\text{range}}^{\min} = \max\{\sqrt{c^2\kappa_{\text{WF}}\Gamma_{\text{eff}}e^{\sigma^2_\phi}}, c\sqrt{\sigma^2_\phi}/(2\pi f_c)\}$ mandates co-optimization of hardware quality ($\Gamma_{\text{eff}} < 0.01$) and oscillator stability ($\sigma^2_\phi < 10^{-2}$) for sub-millimeter performance.

\subsection{Impact of Network Geometry and Interference}

\begin{figure*}[!t]
    \centering
    \includegraphics[width=\textwidth]{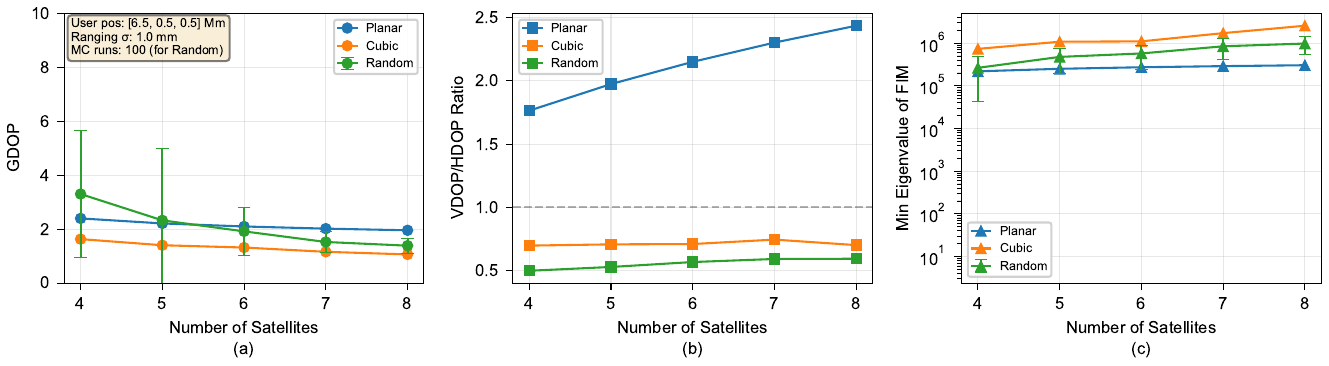}
\caption{Impact of network geometry on sensing performance. (a) GDOP vs. satellite count. (b) VDOP/HDOP ratio. (c) Minimum EFIM eigenvalue (observability strength).}

    \label{fig:geometry}
\end{figure*}

\begin{figure*}[!t]
    \centering
    \includegraphics[width=0.9\textwidth]{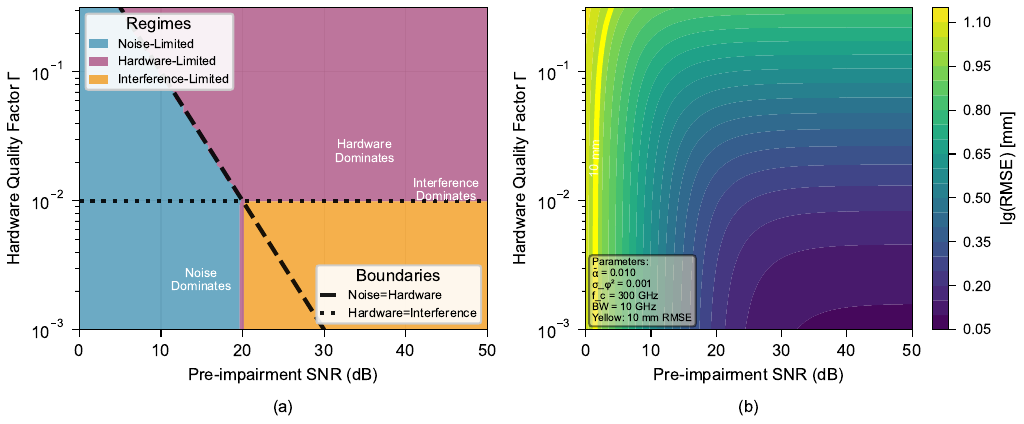}
\caption{Performance regime analysis. (a) Dominant limitation regimes (noise, hardware, or interference). (b) Ranging RMSE (mm) vs. SNR and hardware quality factor $\Gamma$.}

    \label{fig:regime_map}
\end{figure*}

Network performance emerges from the interplay between geometric configuration and co-channel interference, as quantified in Figs.~\ref{fig:geometry} and \ref{fig:regime_map}.

Fig.~\ref{fig:geometry} reveals geometry's critical role through GDOP analysis. Cubic configurations achieve superior performance (GDOP = 1.06 at $N_v=8$) compared to planar arrangements (1.96), with balanced 3D observability (VDOP/HDOP = 0.7 vs. 2.4). The minimum EFIM eigenvalue confirms cubic superiority in the weakest observability direction. Diminishing returns beyond $N_v=6$ (only 7-19\% GDOP improvement to $N_v=8$) suggest optimal constellation sizing.

Fig.~\ref{fig:regime_map} delineates three operational regimes through the dominant SINR limitation: noise-limited (SNR $<$ 20 dB), hardware-limited (transition at $\text{SNR}_0 = 1/\Gamma_{\text{eff}}$), and interference-limited ($\tilde{\alpha} > \Gamma_{\text{eff}}$). The 1.1 mm RMSE contour reveals $\Delta\log(\Gamma) \times \Delta\text{SNR}_{\text{dB}} = \text{const}$—10× hardware improvement equals 10 dB power increase. Dense networks with $\tilde{\alpha} > 0.01$ face an interference wall requiring active mitigation.

These results establish that sub-centimeter THz LEO-ISAC precision demands simultaneous optimization of: (i) 3D geometric diversity, (ii) hardware quality $\Gamma_{\text{eff}} < 0.01$, and (iii) interference management maintaining $\tilde{\alpha} < \Gamma_{\text{eff}}$.

\subsection{Advanced Cooperative Effects: The Role of Noise Correlation}

\begin{figure*}[!t]
    \centering
    \includegraphics[width=\textwidth]{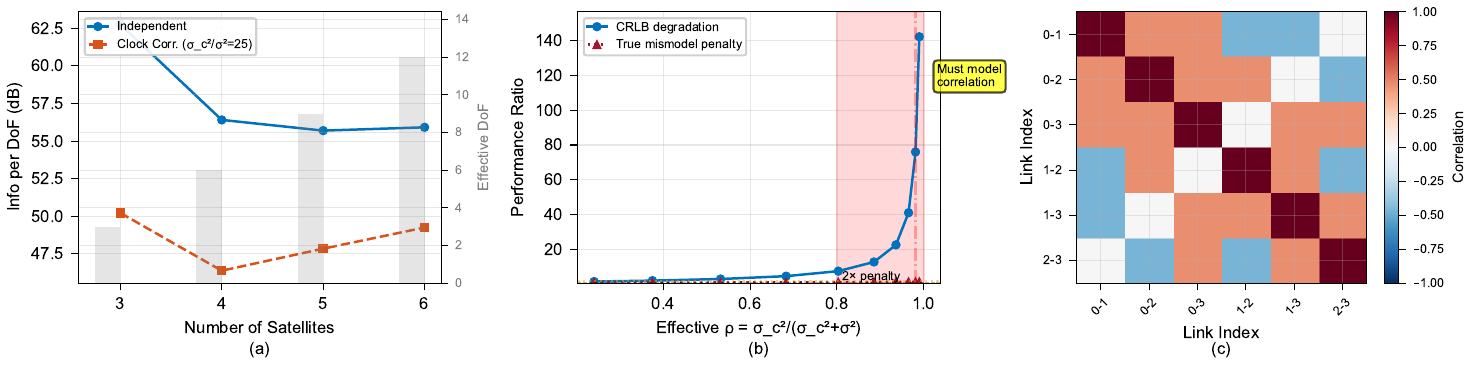}
\caption{Impact of clock-induced noise correlation. (a) Information per DoF vs. satellite count. (b) CRLB degradation and mismodeling penalty vs. correlation. (c) Signed correlation matrix structure.}

    \label{fig:correlation}
\end{figure*}

Clock synchronization architectures   alter network information dynamics through induced measurement correlations. Fig.~\ref{fig:correlation} reveals how statistically dependent noise—traditionally viewed as detrimental—can enhance cooperative sensing through common-mode rejection.

Fig.~\ref{fig:correlation} reveals counterintuitive benefits of noise correlation. While shared timing references ($\mathbf{C}_n = \sigma^2 \mathbf{I} + \sigma^2_c \mathbf{S}\mathbf{S}^T$) reduce information per degree of freedom by 6.7 dB for 6 satellites, actual positioning performance improves through common-mode rejection, maintaining near-unity CRLB degradation at $\rho = 0.9$.

However, mismodeling correlation as independence causes  severe degradation: 53\% performance loss at $\rho > 0.75$, reaching 1.52× penalty at extreme correlation. The shaded region ($\rho > 0.8$) highlights the necessity of correlation modeling, as the true variance $\text{Cov}_{\text{true}} = (\mathbf{H}^T \mathbf{W} \mathbf{H})^{-1} \mathbf{H}^T \mathbf{W} \mathbf{C}_n \mathbf{W} \mathbf{H} (\mathbf{H}^T \mathbf{W} \mathbf{H})^{-1}$ grows unboundedly as $\rho \to 1$.

\begin{figure*}[!t]
    \centering
    \includegraphics[width=0.9\textwidth]{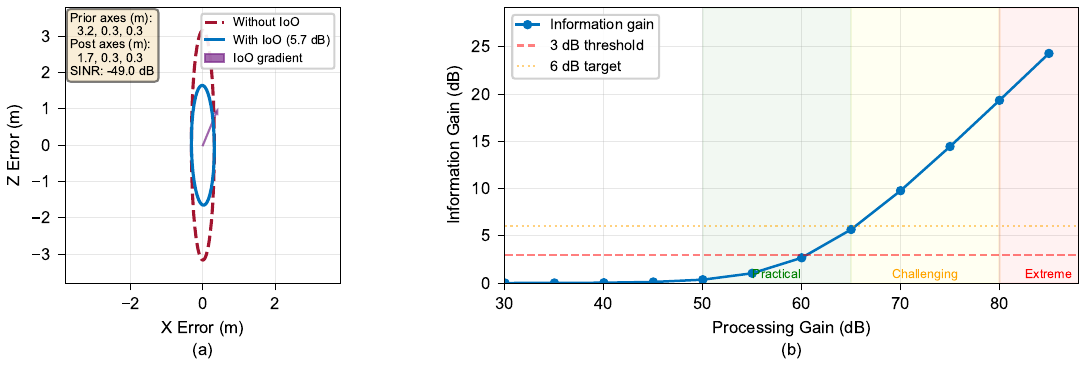}
\caption{Opportunistic sensing (IoO) gain. (a) Error ellipsoid reduction in a geometrically challenged scenario. (b) Information gain vs. processing gain, showing utility thresholds.}

    \label{fig:ioo_gain}
\end{figure*}

\subsection{Quantifying the ISAC Gain via Opportunistic Sensing}

Fig.~\ref{fig:ioo_gain} demonstrates the potential of converting interference into sensing resources through a geometrically compromised scenario where opportunistic sensing provides critical observability.

\subsubsection{Opportunistic Gain in Geometrically-Challenged Scenarios}

Three nearly coplanar satellites face severe vertical observability deficiency, producing error ellipsoid semi-axes of [3.2, 0.3, 0.3] meters---a 10:1 elongation indicating near-singular geometry (Fig.~\ref{fig:ioo_gain}a, dashed red). Traditional solutions require costly additional satellites or orbital maneuvers.

An interfering transmitter contributes orthogonal Fisher Information:
\begin{equation}
\mathbf{J}_{\text{IoO}} = \frac{1}{\sigma^2_{R_b}} \nabla_{\mathbf{p}} R_b (\nabla_{\mathbf{p}} R_b)^T
\end{equation}
Information fusion $\mathbf{J}_{\text{total}} = \mathbf{J}_{\text{prior}} + \mathbf{J}_{\text{IoO}}$ reduces the major axis by 47.8\% to [1.7, 0.3, 0.3] meters (solid blue), achieving 5.67 dB information gain despite -49 dB echo SINR.

\subsubsection{Processing Gain Requirements}

Fig.~\ref{fig:ioo_gain}(b) identifies three operational regions with 3 dB and 6 dB utility thresholds clearly marked:
\begin{itemize}
\item \textbf{Practical (50-65 dB):} Achievable with spread-spectrum and 10-100 ms coherent integration. The 3 dB utility threshold occurs at 65 dB, while the 6 dB gain requires 70 dB processing gain, necessitating 50 dBi×2 antenna gains and typical power consumption of 100-200 W.
\item \textbf{Challenging (65-80 dB):} Requires synthetic aperture or AI-enhanced processing for 6 dB target gain, with integration times approaching 1 second.
\item \textbf{Extreme ($>$80 dB):} Approaches theoretical limits with diminishing returns.
\end{itemize}

The processing gain threshold is determined by\footnote{Derived from bistatic radar equation with $L_{\text{bistatic}} \propto R_{\text{tx}}^2 R_{\text{rx}}^2$ path loss and weak axis alignment of $\mathbf{J}_{\text{prior}}$}:
\begin{equation}
\text{PG}_{\text{threshold}} = 10\log_{10}\left(\frac{L_{\text{bistatic}} \cdot N_{\text{eff}}}{\sigma_b \cdot G_T \cdot G_R}\right)
\end{equation}

Fig.~\ref{fig:ioo_gain}(b) reveals processing gain thresholds: 65 dB for 3 dB utility (achievable with spread-spectrum), 70 dB for 6 dB gain (requiring synthetic aperture), determined by $\text{PG}_{\text{threshold}} = 10\log_{10}(L_{\text{bistatic}} \cdot N_{\text{eff}}/(\sigma_b \cdot G_T \cdot G_R))$\footnote{Bistatic radar equation with $L_{\text{bistatic}} \propto R_{\text{tx}}^2 R_{\text{rx}}^2$.}.

\subsubsection{The ISAC Paradigm Shift}

The crossover where opportunistic gain exceeds interference penalty occurs when:
\begin{equation}
\frac{\text{tr}(\mathbf{J}_{\text{IoO}})}{\lambda_{\min}(\mathbf{J}_{\text{prior}})} > \frac{\alpha_{\ell m}}{\text{SINR}_{\text{comm}}}
\end{equation}
achieving substantial benefit in our scenario, with the information gain of 5.67 dB outweighing the interference penalty. This validates the central thesis: THz LEO networks can transform interference from liability to asset. The demonstrated 5.67 dB gain with realistic processing highlights the value of opportunistic sensing for next-generation satellite networks, offering a strong alternative to traditional interference mitigation paradigms.

\section{Conclusion}

The theoretical framework developed here reveals three fundamental design principles for THz LEO-ISAC networks. First, hardware physics imposes ultimate performance boundaries—ranging precision saturates at $\sigma_{\text{min}} = \max\{\sqrt{c^2\kappa_{\text{WF}}\Gamma_{\text{eff}}e^{\sigma^2_\phi}}, c\sqrt{\sigma^2_\phi}/(2\pi f_c)\}$ regardless of transmit power, with sub-millimeter accuracy demanding co-optimized hardware quality ($\Gamma_{\text{eff}}<0.01$) and oscillator stability ($\sigma^2_\phi<10^{-2}$). Second, network geometry dominates sensing precision—cubic configurations achieve 46\% lower GDOP than planar arrangements with identical satellite counts, establishing three-dimensional diversity as essential rather than optional. Third, interference exhibits fundamental duality—the same signals degrading communication performance provide 5.67 dB sensing gains through opportunistic bistatic measurements when processing gain exceeds 65 dB, transforming traditional mitigation strategies into dynamic resource allocation problems.

This theoretical foundation enables critical advances in next-generation satellite systems. The framework directly supports development of: (i) distributed resource allocation algorithms that dynamically balance the interference-sensing tradeoff quantified by the $\alpha_{\ell m}$ coefficients and opportunistic Fisher information gains; (ii) robust estimation techniques handling non-Gaussian perturbations and atmospheric scintillation effects below 500 km altitude through extended impairment models; and (iii) machine learning approaches that approach the derived N-CRLB bounds through real-time network optimization. The transformation of interference from constraint to resource—demonstrated through 5.67 dB opportunistic sensing gains—  challenges and expands traditional constellation design principles, suggesting that future THz LEO networks should intentionally engineer controlled interference patterns rather than minimize them. As mega-constellations scale toward tens of thousands of nodes, the performance boundaries established here provide both theoretical limits and practical pathways for realizing the full potential of networked ISAC systems.

\bibliographystyle{IEEEtran}
\bibliography{references}

\vspace{-1.2cm} 

\begin{IEEEbiography}
[{\includegraphics[width=1in,height=1.25in,clip]{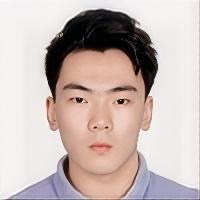}}]{Haofan Dong(hd489@cam.ac.uk)}
 is a Ph.D. student in the Internet of Everything (IoE) Group, Department of Engineering, University of Cambridge, UK. He received his MRes from CEPS CDT based in UCL in 2023. His research interests include integrated sensing and communication (ISAC), space communications, and THz communications.
\end{IEEEbiography}
\vspace{-1.2cm} 

\begin{IEEEbiography}[{\includegraphics[width=1in,height=1.25in,clip]{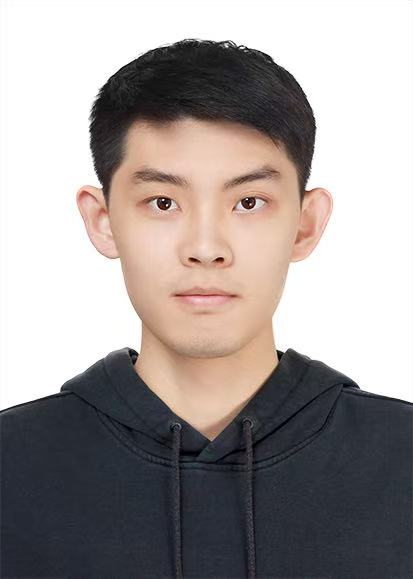}}]{Houtianfu Wang(hw680@cam.ac.uk)}
Houtianfu Wang's current research interests include space communications, ISCC, and semantic communications. He joined University of Cambridge as a Ph.D. student in 2024. He received M.S. degree in Electrical and Computer Engineering from University of California, San Diego in 2024, and B.S. degree in Electrical Engineering from The Ohio State University in 2022. He is currently a member in Internet of Everything group, under the supervision of Prof. Akan.
\end{IEEEbiography}
\vspace{-1.2cm} 

\begin{IEEEbiography}[{\includegraphics[width=1in,height=1.25in,clip]{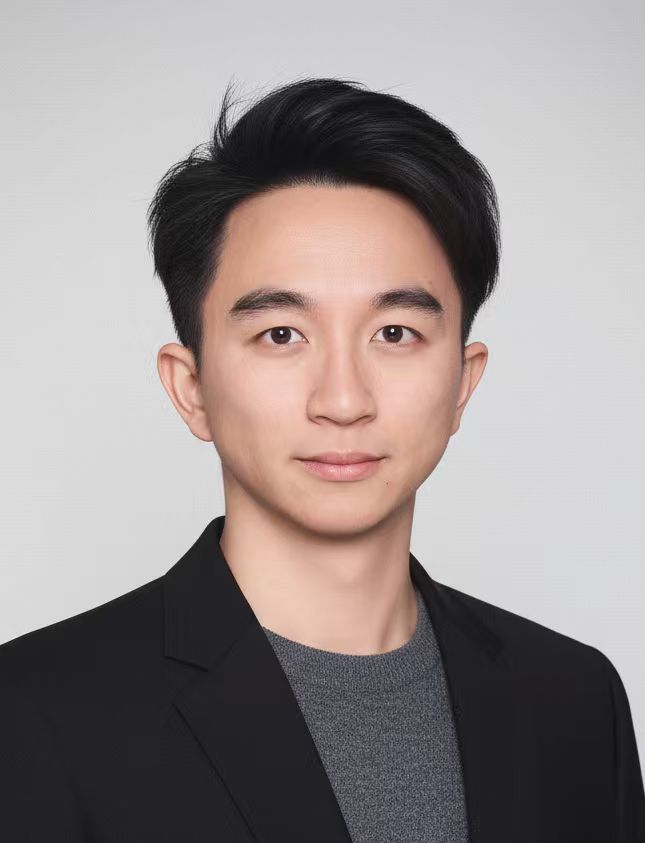}}]{Hanlin Cai(hc663@cam.ac.uk)} received the B.Sc degree from the Engineering College, National University of Ireland, Maynooth in 2024, and the M.Phil. degree from the Department of Engineering, University of Cambridge in 2025, where he is currently pursuing the Ph.D. degree. His research interests include wireless communication, wireless federated learning, and the Internet of Agents.
\end{IEEEbiography}
\vspace{-1.2cm} 

\begin{IEEEbiography}[{\includegraphics[width=1in,height=1.25in,clip]{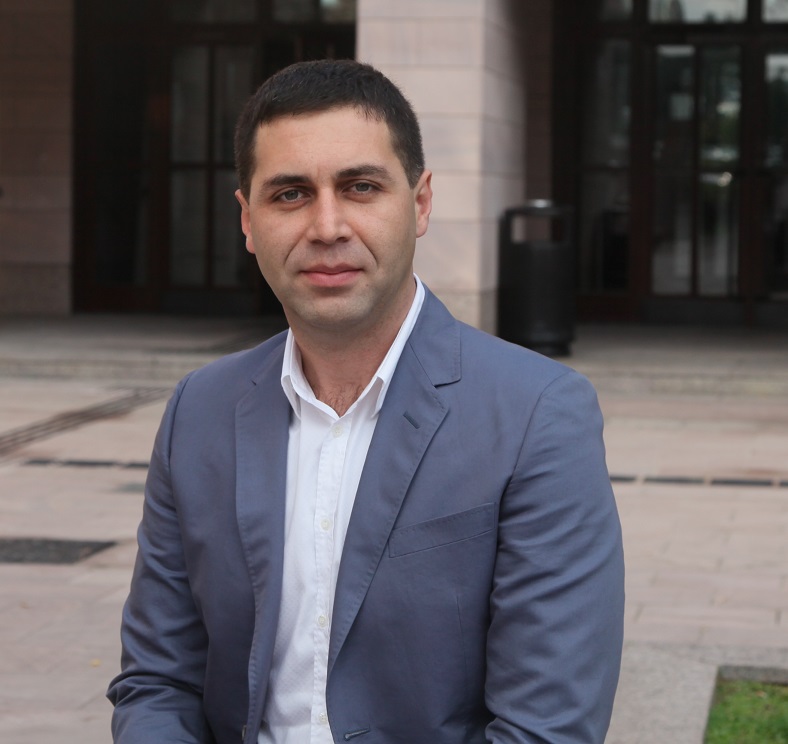}}]{Ozgur B. Akan(oba21@cam.ac.uk) }
received his Ph.D. degree from the School of Electrical and Computer Engineering, Georgia Institute of Technology, Atlanta, in 2004. He is currently the Head of the Internet of Everything (IoE) Group, Department of Engineering, University of Cambridge, and the Director of the Centre for NeXt-Generation Communications (CXC), Koç University. His research interests include wireless, nano-, and molecular communications, and the Internet of Everything.
\end{IEEEbiography}
\vspace{-1.2cm} 

\end{document}